\documentclass[aps, prb, twocolumn, superscriptaddress]{revtex4-2}
\usepackage{graphicx}
\usepackage{dcolumn}
\usepackage{bm}
\usepackage{amsmath}

\begin{document}

\title{Decoupling interface and thickness effects on hydrogen absorption in V/MgO: experiments and DFT}

\author{Qiuxiang Zhang}
\affiliation{School of Electronic Technology and Engineering, Shanghai Technical Institute of Electronics and Information, Shanghai, 201411, China}

\author{Yan Zhu}
\affiliation{School of Electronic Technology and Engineering, Shanghai Technical Institute of Electronics and Information, Shanghai, 201411, China}

\author{Xiaofang Peng}
\affiliation{School of Electronic Technology and Engineering, Shanghai Technical Institute of Electronics and Information, Shanghai, 201411, China}

\author{Weiguang Yang}
\affiliation{School of Materials Science and Engineering, Shanghai University, Shanghai, 200444, China}

\author{Yuping Le}
\affiliation{School of Electronic Technology and Engineering, Shanghai Technical Institute of Electronics and Information, Shanghai, 201411, China}

\author{Xiao Xin}
\affiliation{School of Electronic Technology and Engineering, Shanghai Technical Institute of Electronics and Information, Shanghai, 201411, China}

\date{November 7, 2025}

\begin{abstract}
We report combined experimental and first-principles investigations of hydrogen absorption in epitaxial vanadium films on MgO(001) with nominal thicknesses of 10\,nm and 50\,nm. In-situ optical transmission and four-probe resistance isotherms show that the 50\,nm film reproduces bulk-like behavior with a clear first-order $\alpha$--$\beta$ hydride transition, the formation enthalpy and entropy gradually decrease with increasing hydrogen concentration. The 10\,nm film, by contrast, displays continuous uptake without plateaus, with formation enthalpies $\Delta H$ that are relatively close in magnitude to the 50\,nm film (both exhibiting exothermic behavior in the range of approximately $-0.5$ to $-0.3$\,eV/H), but with a more negative entropy change $\Delta S$ (larger $|\Delta S|$) indicating reduced configurational freedom for hydrogen in the ultrathin limit; the critical temperature for phase coexistence is suppressed below 400 K. Density functional theory calculations on MgO/V superlattices (V$_n$/(MgO)$_n$, $n=3,5,7$) reveal pronounced V\,3d--O\,2p hybridization and interfacial charge redistribution that weaken hydrogen binding near the interface and recover toward bulk values with increasing V thickness. These results indicate that interfacial electronic structure, in addition to finite-size energetics, governs hydride stability in ultrathin V films and that layer-thickness and interface engineering can tune reversible hydrogen uptake.
\end{abstract}

\maketitle

\section{Introduction}

Hydrogen interaction with transition metals governs key phenomena in energy storage, embrittlement and surface chemistry; bulk metal--hydrogen thermodynamics are well established \cite{fukai,Pundt2006,Luo1990,veleckis1969}, while confinement and interfaces introduce additional energetic and entropic contributions that can qualitatively change phase behaviour \cite{Li2013,XinThesis}. Understanding these contributions is essential for both fundamental science and applications that require reversible hydrogen uptake or mitigation of hydrogen-induced damage \cite{Pundt2006}.

Vanadium is a prototypical hydride-forming transition metal: in bulk and in sufficiently thick films hydrogen absorption produces a first-order $\alpha$--$\beta$ transition that is routinely characterized by optical and transport probes \cite{jan2010,pryde1971,palsson2010}. Recent thin-film experiments demonstrate a strong thickness dependence of the critical temperature and of isotherm character, with pronounced finite-size suppression of phase coexistence in the nanometer regime \cite{Xin2014,XinThesis,Huang1994}. These trends parallel finite-size effects observed in other ordering phenomena and motivate a systematic decoupling of interface and size contributions to the hydrogen thermodynamics \cite{Hjorvarsson1991,meded2005}.

Interfaces modify hydrogen behaviour through multiple, concurrent mechanisms: altered coordination and local crystal field, strain and lattice mismatch, interfacial free-energy penalties for phase boundaries, and electronic reconstruction that changes local binding energies \cite{Hjorvarsson1991,meded2005,Li2013}. MgO(001) is a convenient, well-controlled substrate for epitaxial V(001) growth because of favourable lattice matching and chemical stability \cite{isberg1997,palsson2008}. A thin Pd cap is commonly used to catalyze H$_2$ dissociation while minimally perturbing the V response \cite{jan2010,ohrmalm1999,prox,bloch2010,andersson1997a,andersson1997b,olsson2002,olsson2005}; prior experimental studies of caps, alloyed interfaces and multilayers further illustrate how substrate/cap chemistry influences uptake and transport \cite{palsson2010,prox,ohrmalm1999,bloch2010}.

First-principles calculations complement experiment by revealing how interfacial bonding and charge redistribution modify hydrogen binding and the local density of states. Earlier DFT work on MgO/V and related systems reports pronounced V\,3d--O\,2p hybridization and thickness-dependent electronic reconstruction that affect magnetic, transport and chemical properties \cite{palsson2012,palsson2012_natcom,aboud2010}; practical computations typically employ the PBEsol exchange-correlation functional with ultrasoft pseudopotentials generated using the Vanderbilt code (version 7.3.6), including nonlinear core corrections and scalar relativistic effects with careful convergence and ZPE estimates \cite{pbesol,vanderbilt_uspp,Li2013}. In this study we combine temperature- and pressure-dependent in-situ optical transmission and four-probe resistance measurements on epitaxial V films (nominally 10\,nm and 50\,nm) with DFT on V$_n$/(MgO)$_n$ ($n=3,5,7$) superlattices to disentangle interfacial electronic effects from finite-size energetic contributions and to identify the microscopic origin of thickness-dependent hydride destabilization.

\section{Experimental Methods and Results}

\begin{figure*}[htbp]
    \centering
    \includegraphics[width=1\linewidth]{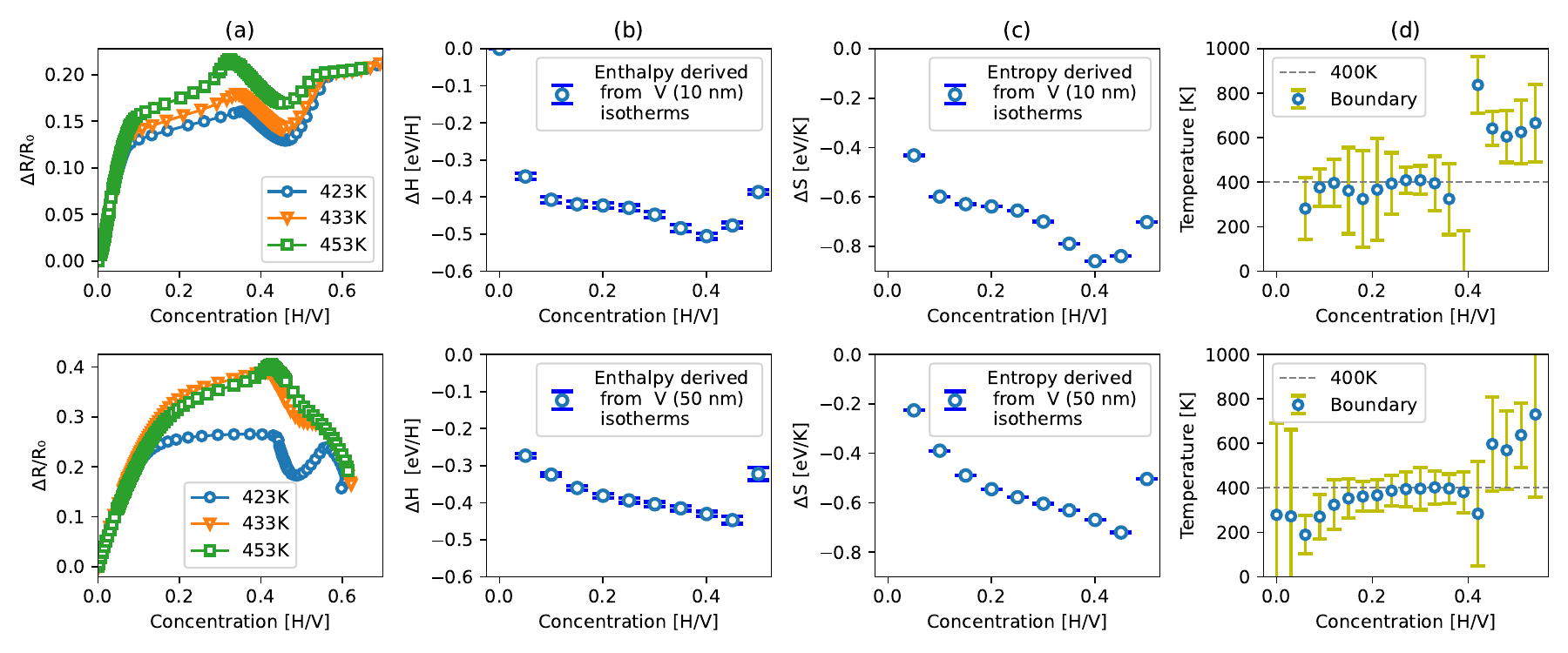}
    \caption{
        Thermodynamic characterization of hydrogen absorption in epitaxial V films on MgO(001): 10\,nm (top row) and 50\,nm (bottom row).
        (a) Normalized resistance change $\Delta R/R_0$ versus hydrogen concentration [H/V]; plateaus in the 50\,nm film mark a first-order $\alpha$--$\beta$ transition, while the 10\,nm film shows continuous uptake.
        (b) Hydride formation enthalpy $\Delta H$ from van't Hoff analysis. The $\Delta H$ values for the 10\,nm and 50\,nm films are relatively close in magnitude, both exhibiting exothermic behavior in the range of approximately $-0.5$ to $-0.3$\,eV/H, suggesting modest thickness dependence of the binding energetics.
        (c) Entropy change $\Delta S$ as a function of concentration: the 10\,nm film displays a more negative $\Delta S$ (larger $|\Delta S|$) than the 50\,nm film, indicating restricted configurational freedom or less degrees of freedom for hydrogen in the ultrathin limit.
        (d) Calculated phase boundaries from $(\partial\mu/\partial c)_T=0$. The critical temperature for $\alpha$--$\beta$ coexistence exceeds $400\,$K for the 50\,nm film but is suppressed below $400\,$K for the 10\,nm film across the measured range.
    }
    \label{fig:thermo}
\end{figure*}

Epitaxial vanadium films (nominal thicknesses 10\,nm and 50\,nm) were grown on single-crystal MgO(001) substrates (10\,mm$\times$10\,mm$\times$0.1\,mm) by DC magnetron sputtering under base pressures $<5\times10^{-8}$\,Torr, following the procedures of Refs.~\onlinecite{isberg1997,jan2010}. Films were capped \textit{in situ} with 5\,nm Pd to prevent oxidation and to catalyze H$_2$ dissociation; previous studies show that, under our measurement conditions, the Pd cap contributes negligibly to the measured hydride signal of the underlying V layer \cite{jan2010,ohrmalm1999,prox,bloch2010,andersson1997a,andersson1997b}. Film thicknesses and crystalline quality were confirmed by x-ray reflectivity and low-angle x-ray diffraction (not shown), consistent with earlier epitaxial V/MgO reports \cite{palsson2008,palsson2010}.

Hydrogen uptake was measured in a custom UHV-compatible chamber enabling simultaneous, \textit{in situ} optical transmission and four-probe electrical resistance as functions of hydrogen pressure $p$ and temperature $T$. Optical transmission at 639\,nm (LED source, lock-in detection) was converted to atomic concentration $c$ (H/V) using a calibrated Beer--Lambert relation $I/I_0=\exp(-A\,c)$, where the scale factor $A$ was obtained by referencing bulk vanadium hydride data and the 50\,nm film \cite{fukai,jan2010,palsson2010}. Four-probe resistance was recorded to remove contact contributions and reported as normalized change $\Delta R/R_0=[R(c,T)-R(0,T)]/R_0$, providing a sensitive indicator of electronic/structural changes associated with ordering or phase separation \cite{pryde1971,jan2010}.

Isotherms were collected by equilibrating at fixed $T$ while stepping $p$ over the range relevant for the $\alpha$--$\beta$ transition. Equilibration criteria required pressure and optical/resistive signals to be stable within experimental noise for at least 5--10\,min. From the temperature dependence of equilibrium pressures at fixed concentrations we constructed van't Hoff plots to extract formation enthalpies $\Delta H$ and entropies $\Delta S$; uncertainties incorporate calibration errors, temperature stability, and fitting uncertainty (see error bars in figures). Our calibration approach and analysis are consistent with published methodology for thin-film hydrides and multilayers \cite{Luo1990,veleckis1969,prox,ohrmalm1999,olsson2002,olsson2005}.

The combined dataset (optical and transport) offers complementary sensitivity: transmission provides an absolute concentration scale tied to bulk references \cite{fukai,jan2010}, while $\Delta R/R_0$ reveals ordering-related electronic signatures even when macroscopic phase separation is suppressed \cite{pryde1971,palsson2010,bloch2010}. Representative raw isotherms and derived thermodynamic quantities are presented in Fig.~\ref{fig:thermo}. The 50\,nm films reproducibly show pressure--concentration plateaus and a nearly constant, strongly exothermic $\Delta H$, consistent with bulk-like behavior. In contrast, the 10\,nm films exhibit smooth, continuous uptake without plateaus, with a $\Delta H$ that is only modestly less exothermic and lies within a similar range ($\sim -0.5$ to $-0.3$\,eV/H), but displays a more negative entropy change $\Delta S$ (larger $|\Delta S|$), suggesting reduced configurational freedom or less degrees of freedom for hydrogen in the ultrathin limit.

We attribute these thickness-dependent trends to the combined effects of (i) interfacial energetics that penalize phase boundary formation in confined layers and (ii) interfacial electronic reconstruction that modifies local H binding. These interpretations are supported by prior experimental and theoretical work on V-based thin films and capped multilayers \cite{Xin2014,XinThesis,Hjorvarsson1991,meded2005,palsson2012,palsson2012_natcom}. The following section presents DFT calculations that quantify the spatial variation of hydrogen binding and the electronic-structure changes at the MgO/V interface, providing a microscopic basis for the observed experimental thermodynamics.

\section{Computational Methods}

\begin{figure}[htbp]
    \centering
    \includegraphics[width=0.7\linewidth]{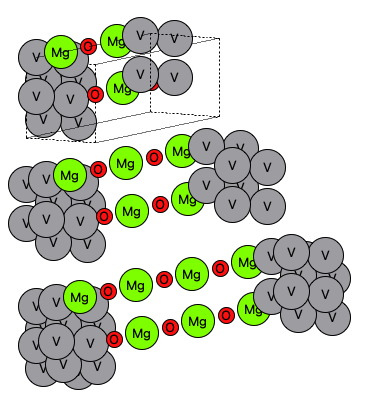}
    \caption{Primitive cell samples of MgO-V superlattice structure used in DFT calculations.}
    \label{fig:structure}
\end{figure}

First-principles density functional theory (DFT) calculations were performed using the Quantum ESPRESSO package with the PBEsol exchange-correlation functional. To model epitaxial V films on MgO(001), we constructed superlattices of V$_n$/(MgO)$_n$ ($n=3,5,7$) with periodic boundary conditions, as illustrated in Fig.~\ref{fig:structure}. The in-plane lattice constant was fixed to the optimized MgO value (4.21\,\AA) to simulate epitaxial strain, while the out-of-plane lattice constant and atomic positions were fully relaxed until forces were below 0.01\,eV/\AA.

Ultrasoft pseudopotentials were employed with plane-wave cutoffs of 50--60\,Ry (charge-density cutoff 400--600\,Ry), determined through convergence testing. The vanadium pseudopotential was specifically generated using the Vanderbilt code (version 7.3.6) with PBEsol exchange-correlation functional, including nonlinear core correction and scalar relativistic treatment. This pseudopotential features 13 valence electrons with a maximum angular momentum component of $l=2$ (d-states) and utilizes 4 wavefunctions with 6 projectors. Brillouin-zone integration used $\Gamma$-centered Monkhorst--Pack meshes converged to $\lesssim$1\,meV/atom (typical grids: $4\times4\times1$ for $n=3$, $6\times6\times1$--$2$ for $n=5,7$). A small cold smearing was applied for metallic systems, and PDOS were generated with 0.05\,eV Gaussian broadening. These computational parameters follow established protocols for V/MgO systems and thin-film hydrides~\cite{palsson2008,palsson2010,palsson2012,palsson2012_natcom,aboud2010,prox}.

Hydrogen binding energies were evaluated by placing H in symmetry-allowed interstitial sites (tetrahedral and octahedral) at varying distances from the MgO/V interface. Binding energies per H are defined as
\[
E_b = E[\mathrm{V/MgO+H}] - E[\mathrm{V/MgO}] - \tfrac{1}{2}E[\mathrm{H_2}],
\]
with the H$_2$ reference computed in a large cubic cell. Selected configurations were corrected for zero-point energy (ZPE) using finite-displacement phonon calculations at Gamma; the ZPE procedure and associated uncertainties follow prior work on thin-film hydrides and multilayers \cite{Luo1990,fukai,prox,ohrmalm1999}.

Electronic-structure analysis includes projected density of states (PDOS) onto atomic orbitals, site-resolved density differences, and Bader charge partitioning to quantify interfacial charge transfer \cite{palsson2012,aboud2010}. Where relevant we report site-resolved shifts in V\,3d and O\,2p spectral weight and correlate these with spatially resolved H binding energies. The overall computational strategy follows protocols employed in related experimental/theoretical studies of V thin films, caps and multilayers \cite{jan2010,isberg1997,bloch2010,ohrmalm1999,andersson1997a,andersson1997b,olsson2002,olsson2005,Huang1994,Li2013,meded2005,fukai}.

All energies are reported relative to the Fermi level (set to 0\,eV) and include the stated ZPE corrections where specified. The computational results presented below were used to construct binding-energy maps and to interpret the experimentally observed thickness dependence of hydride thermodynamics.

\section{Electronic Structure Analysis}

\begin{figure}[htbp]
    \centering
    \includegraphics[width=1\linewidth]{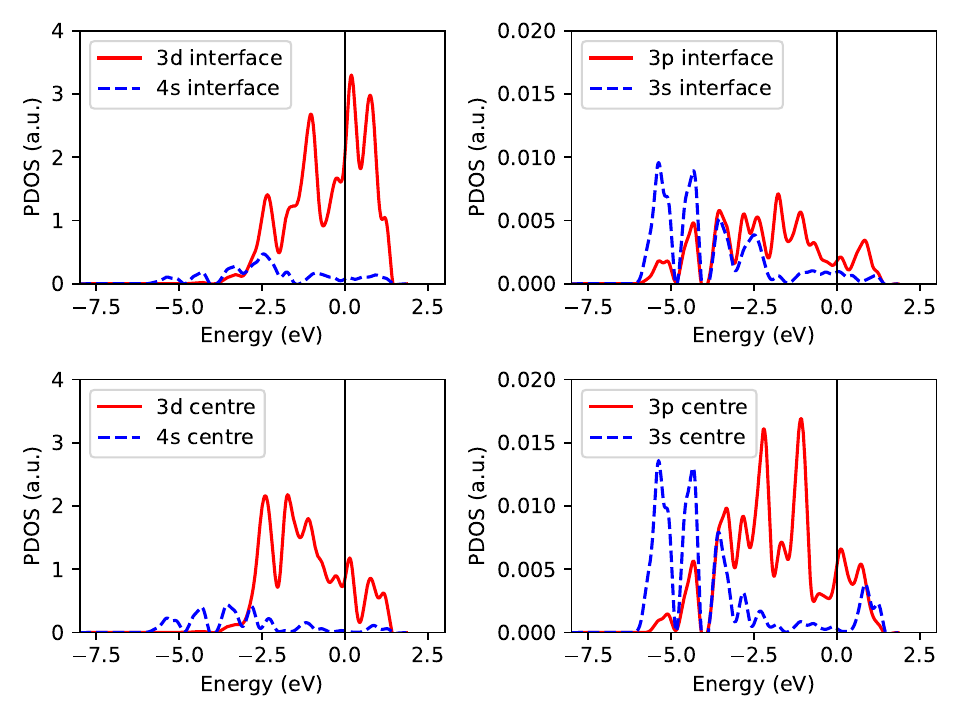}
    \caption{Projected density of states (PDOS) for V $3d$ and $4s$ orbitals at interface (top panels) and center (bottom panels) sites, and V $3p$ and $3s$ orbitals in V$_3$(MgO)$_3$. The Fermi level is set at 0 eV.}
    \label{fig:pdos_v3}
\end{figure}

Calculations of V$_3$(MgO)$_3$, V$_5$(MgO)$_5$, and V$_7$(MgO)$_7$ reveal a consistent and dramatic modification of the electronic structure for V atoms at the MgO interface compared to those in the center (bulk-like region) \cite{palsson2012}. The PDOS of central V atoms closely resembles bulk vanadium, dominated by V-3d states near $E_F$. In contrast, interface V atoms exhibit strong hybridization between V-3d and O-2p states, indicating a covalent interaction responsible for interfacial adhesion and charge redistribution \cite{aboud2010}.

The electronic structure of the V$_3$(MgO)$_3$ system was investigated using the PBEsol exchange-correlation functional with the Vanderbilt-generated ultrasoft pseudopotential. A $4 \times 4 \times 1$ Monkhorst-Pack \textbf{k}-point grid was used for Brillouin zone integration. The tetragonal crystal structure (space group \textit{P4/mmm}, \texttt{ibrav=6}) with lattice parameters $a = 11.259$ a.u. and $c/a = 3.138$ was adopted. As shown in Fig.~\ref{fig:pdos_v3}, the V $3d$ orbitals dominate the states crossing the Fermi level ($E_F$), indicating their crucial role. The contribution from V $4s$ orbitals is significantly smaller but non-negligible near $E_F$, while V $3p$ and $3s$ orbitals contribute primarily to the core states. The distinct difference in PDOS between interface and center V atoms highlights a site-dependent electronic environment. The enhanced V-3d spectral weight near $E_F$ at the interface (compared to central sites) arises from reduced coordination and band narrowing effects, consistent with strong V-3d/O-2p hybridization observed in thicker superlattices.

\begin{figure}[htbp]
    \centering
    \includegraphics[width=1\linewidth]{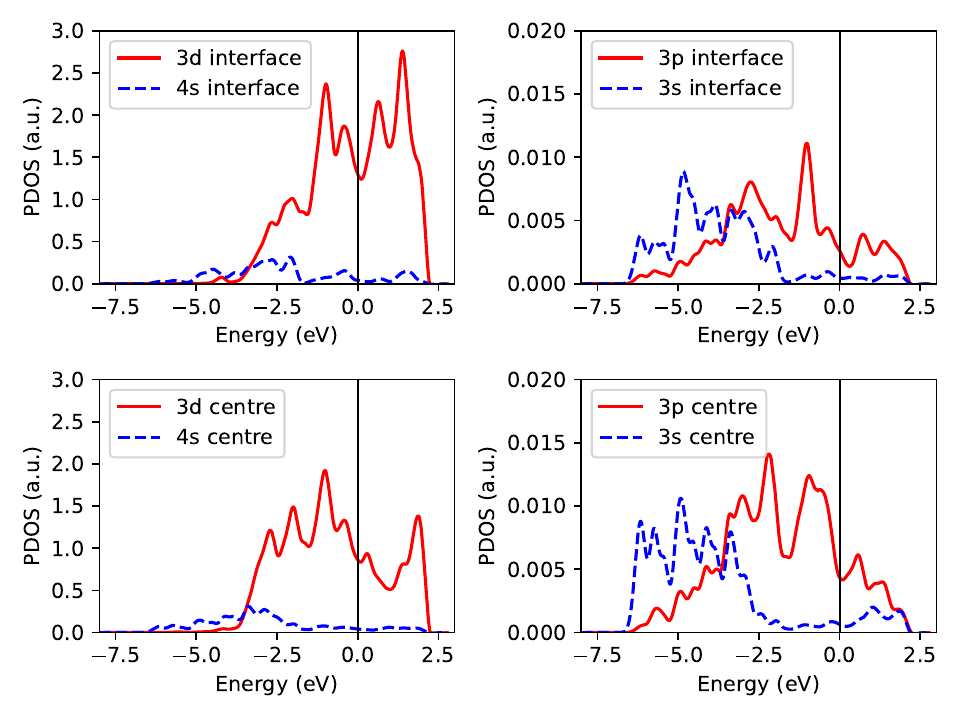}
    \caption{Partial density of states of V $3d$, $4s$, $3p$ and $3s$ orbitals at interface and center sites in V$_5$(MgO)$_5$. The Fermi level is set at 0 eV.}
    \label{fig:pdos_v5}
\end{figure}

The electronic structure of the V$_5$(MgO)$_5$ system, shown in Fig.~\ref{fig:pdos_v5}, is primarily governed by V-3d and O-2p states. A critical observation is the pronounced enhancement of V-3d spectral weight near $E_F$ for interface atoms, which reflects the fundamental physics of reduced coordination at the interface. This electronic reconstruction is characterized by narrower band widths and higher PDOS amplitudes due to enhanced electron localization. The O-2p states hybridize strongly with V-3d states in the energy range of -5 to -2 eV, forming covalent bonds that contribute to interfacial adhesion.

\begin{figure}[htbp]
    \centering
    \includegraphics[width=1\linewidth]{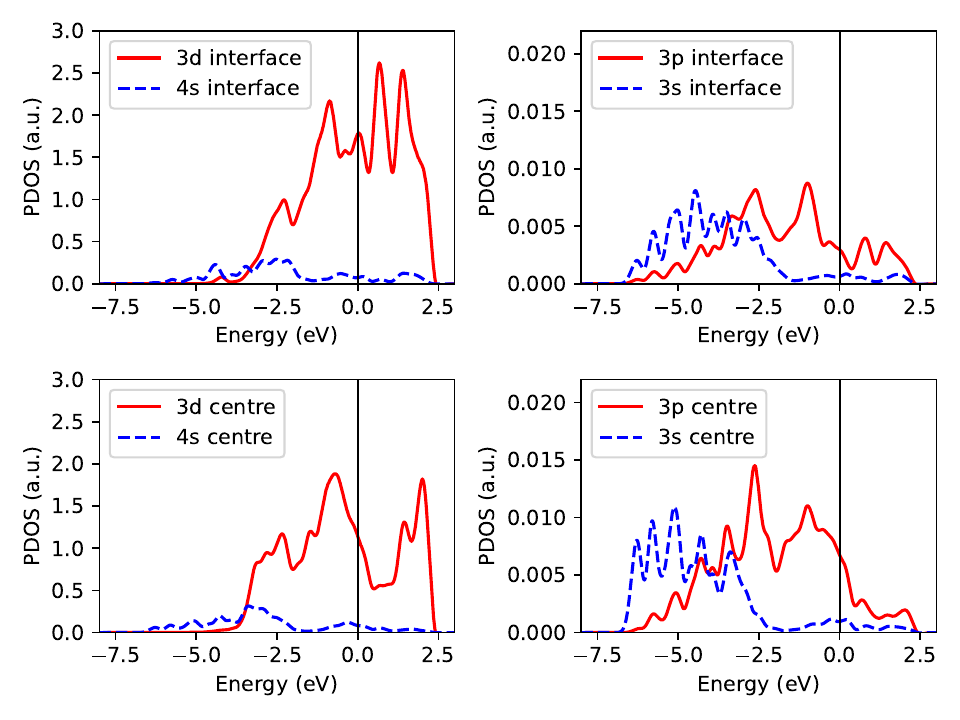}
    \caption{Partial density of states of V $3d$, $4s$, $3p$ and $3s$ orbitals at interface and center sites in V$_7$(MgO)$_7$. The Fermi level is set at 0 eV.}
    \label{fig:pdos_v7}
\end{figure}

The electronic structure of V$_7$(MgO)$_7$ (Fig.~\ref{fig:pdos_v7}) is dominated by V-3d states, with interface V layers exhibiting characteristically sharper features and enhanced spectral weight near $E_F$ compared to central V layers. Notably, a systematic thickness-dependent evolution is observed in the central V layers: as the slab thickness increases from V$_3$(MgO)$_3$ to V$_5$(MgO)$_5$ and finally to V$_7$(MgO)$_7$, the V-3d spectral features in the central layers progressively sharpen and develop enhanced coherence near $E_F$, gradually converging toward the characteristic electronic structure of bulk vanadium. This interfacial electronic modification characterized by both the modified crystal field at V sites and increased covalent bonding with O—provides a direct electronic mechanism for the observed thickness-dependent hydrogen thermodynamics.

The extent of the V-3d/O-2p hybridization is sensitive to superlattice periodicity. This interface-induced electronic changes is most pronounced in V$_3$(MgO)$_3$ (thinnest V layer), where the interface-to-volume ratio is highest, and gradually weakens as the V layer thickness increases (V$_5$(MgO)$_5$, V$_7$(MgO)$_7$), converging towards the bulk-like PDOS. This trend is consistent with the enhanced V-3d spectral weight near $E_F$ at the interface, which is most prominent in the thinnest superlattice (Fig.~\ref{fig:pdos_v3}) and diminishes with increasing V thickness (Figs.~\ref{fig:pdos_v5} and \ref{fig:pdos_v7}). This tunable interfacial electronic structure has profound implications for hydrogen absorption thermodynamics, where interfacial sites govern the local hydrogen chemical potential \cite{palsson2012_natcom}.

\section{Results and Discussion}

We compare hydrogen absorption thermodynamics in epitaxial V films of nominal thickness 50\,nm and 10\,nm. Experimentally (Fig.~\ref{fig:thermo}), the 50\,nm films exhibit well-defined pressure--concentration plateaus and abrupt features in $\Delta R/R_0$, consistent with a bulk-like, first-order $\alpha$--$\beta$ hydride transition. By contrast, the 10\,nm films show smooth, continuous uptake with no plateau, indicating suppression of macroscopic phase coexistence in the ultrathin limit \cite{Xin2014,XinThesis,Huang1994,Li2013}.

Van't Hoff analysis reveals that while the formation enthalpy $\Delta H$ values for the 10\,nm and 50\,nm films are relatively close in magnitude (both exhibiting exothermic behavior in the range of approximately $-0.5$ to $-0.3$\,eV/H), the entropy change $\Delta S$ displays a striking thickness dependence. The 10\,nm film shows a more negative $\Delta S$ (larger $|\Delta S|$) than the 50\,nm film, indicating restricted configurational freedom or less degrees of freedom for hydrogen in the ultrathin regime. This entropy restriction, rather than enthalpy destabilization, governs the suppression of phase coexistence in thin films.

DFT calculations on V$_n$/(MgO)$_n$ ($n=3,5,7$) provide a microscopic basis for these observations through systematic analysis of electronic structure evolution with thickness. The PDOS analysis reveals a dual thickness-dependent phenomenon: (i) interface V layers consistently exhibit sharper spectral features and enhanced V-3d spectral weight near $E_F$ compared to central layers across all thicknesses, and (ii) the central V layers show progressive evolution toward bulk-like electronic structure as thickness increases from V$_3$(MgO)$_3$ to V$_5$(MgO)$_5$ to V$_7$(MgO)$_7$, with V-3d states becoming increasingly coherent and bulk-characteristic. This systematic central-layer evolution directly correlates with the experimental observation that thicker films recover bulk-like phase behavior.

The interfacial electronic reconstruction characterized by modified crystal field at V sites and increased V--O covalent bonding—creates a spatial gradient in hydrogen binding energetics. Hydrogen binding is strongest in the central, bulk-like regions and progressively weakens toward the MgO interface due to charge redistribution and altered local coordination. In ultrathin films, the dominance of interfacial sites relative to bulk-like central sites creates an average binding environment that favors continuous solid-solution behavior rather than phase separation.

Two complementary mechanisms explain the observed thickness-dependent thermodynamics. First, finite-size effects limit the spatial extent available for phase boundary formation, with the energetic penalty for creating $\alpha$--$\beta$ interfaces becoming prohibitive below a critical thickness \cite{Hjorvarsson1991,meded2005,Xin2014}. Second, electronic reconstruction at the MgO/V interface modifies the local density of states near $E_F$, reducing the electronic driving force for ordering transitions that characterize first-order hydride formation \cite{palsson2012,palsson2012_natcom,aboud2010,prox}. The restricted entropy observed in thin films arises from this electronic modification, which creates a distribution of local hydrogen environments with varying binding energies, effectively decreasing the configurational entropy of hydrogen occupation.

Our conclusions are robust against methodological variations. We employed the PBEsol functional with ultrasoft pseudopotentials generated using the Vanderbilt approach, and rigorously performed convergence tests for plane-wave cutoff energy, k-point sampling, and zero-point energy (ZPE) in accordance with established protocols \cite{vanderbilt_uspp,pbesol,fukai}. Taken together, the combined experimental and computational evidence supports a unified picture in which both geometric confinement (finite-size effects) and interfacial electronic structure govern hydride stability in ultrathin vanadium films \cite{Li2013,Huang1994,prox}.

Practically, the results suggest strategies to recover or tune hydride thermodynamics: increasing V thickness, inserting chemically inert spacer layers, or modifying substrate/cap chemistry to reduce V--O hybridization should shift the local H chemical potential toward bulk values and restore phase coexistence. Conversely, deliberate interface design provides a route to suppress hydride formation where that is desirable for device stability or functionality \cite{palsson2012_natcom,jan2010,meded2005}.

Finally, we note that our findings connect to broader metal--hydrogen phenomenology: the observed enthalpy and entropy shifts are consistent with bulk solution energetics reported in the literature and with thin-film/multilayer observations of cap and substrate effects \cite{fukai,Pundt2006,Luo1990,prox,ohrmalm1999}. This cross-validation underscores the generality of interfacial electronic control of hydrogen thermodynamics in confined metallic systems.

\end{document}